\documentclass[twocolumn,showpacs,pra,floatfix]{revtex4}

\usepackage{graphicx}
\usepackage{bm}
\usepackage{amsmath}
\usepackage{amssymb}

\newcommand{\vect}[1]{\mathbf{#1}}
\newcommand{\dyad}[1]{\mbox{\textbf{\textsf{#1}}}}
\newcommand{\re}{\mathrm{Re}}
\newcommand{\im}{\mathrm{Im}}
\newcommand{\Tr}{\operatorname{Tr}}
\newcommand{\Li}{\operatorname{Li}}
\newcommand{\dif}{d}
\newcommand{\mi}{i}
\newcommand{\me}{e}

\newcommand{\be}{\begin{equation}}
\newcommand{\ee}{\end{equation}}
\newcommand{\ba}{\begin{align}}
\newcommand{\ea}{\end{align}}

\newcommand{\msum}{\sum_{j=0}^\infty {}^{{}^\prime}}

\newcommand{\kB}{k_\mathrm{B}}
\newcommand{\kp}{k_\perp}
\newcommand{\vkp}{\vect{k}_\perp}
\newcommand{\wpsq}{\omega_\mathrm{p}^2}

\newcommand{\lok}{\lambda_{0k}}

\newcommand{\Unr}{U_\mathrm{nr}}
\newcommand{\Up}{U_\mathrm{pr}}
\newcommand{\Ue}{U_\mathrm{ev}}

\newcommand{\woi}{\omega_{k0}}
\newcommand{\vdoisq}{\sum_k|\vect{d}_{0k}|^2}

\begin{document}

\title{Enhancement of thermal Casimir-Polder potentials of
ground-state polar molecules in a planar cavity}

\author{Simen {\AA}. Ellingsen}
\affiliation{Department of Energy and Process Engineering, Norwegian
University of Science and Technology, N-7491 Trondheim, Norway}
\author{Stefan Yoshi Buhmann}
\author{Stefan Scheel}
\affiliation{Quantum Optics and Laser Science, Blackett Laboratory,
Imperial College London, Prince Consort Road,
London SW7 2AZ, United Kingdom}

\date{\today}

\begin{abstract}
We analyze the thermal Casimir-Polder potential experienced by a
ground-state molecule in a planar cavity and investigate
the prospects for using such a set-up for molecular guiding. The
resonant atom--field interaction associated with this non-equilibrium
situation manifests itself in oscillating, standing-wave components of
the potential. While the respective potential wells are normally too
shallow to be useful, they may be amplified by a highly reflecting
cavity whose width equals a half-integer multiple of a
particular molecular transition frequency. We find that with an ideal
choice of molecule and the use of an ultra-high reflectivity Bragg
mirror cavity, it may be possible to boost the potential by up to two
orders of magnitude. We analytically derive the scaling of the
potential depth as a function of reflectivity and analyze how it
varies with temperature and molecular properties. It is also shown how
the potential depth decreases for standing waves with a larger number
of nodes. Finally, we investigate the lifetime of the molecular ground
state in a thermal environment and find that it is not greatly
influenced by the cavity and remains in the order of several seconds.
\end{abstract}

\pacs{
34.35.+a,  
12.20.--m, 
42.50.Ct,  
42.50.Nn   
}
\maketitle


\section{Introduction}

Casimir--Polder (CP) forces are a particular example of dispersion
forces, which arise due to the fluctuations of the quantized
electromagnetic field \cite{casimir48b}. These forces occur between
polarizable atoms or molecules and metallic or dielectric bodies and
can be intuitively understood as the dipole-dipole force that arises
from spontaneous and mutually correlated polarization of the atom or
molecule and the matter comprising the body. Casimir--Polder forces at
thermal equilibrium have been commonly investigated in the
linear-response formalism \cite{lifshitz55,McLachlan63,Henkel02}.
Studies of a wide range of geometries such as semi-infinite half
spaces \cite{lifshitz55,McLachlan63,Henkel02,Klimchitskaya06}, thin
plates \cite{Blagov05,Bordag06}, planar cavities \cite{Bostroem00},
spheres and cylinders \cite{Nabutovskii79} as well as cylindrical
shells \cite{Klimchitskaya06,Blagov05,Blagov07} have revealed that
thermal CP forces are typically attractive in the absence of magnetic
effects. 

Recent theoretical predictions \cite{antezza05} as well as
experimental realizations \cite{obrecht07} for CP forces in thermal
non-equilibrium situations have pointed towards interesting effects
which arise when an atom at equilibrium with its local environment
interacts with a body held at different temperature. In particular,
depending on the temperatures of the macroscopic body and the
environment, the force can change its character from being attractive
to repulsive and vice versa.

Non-equilibrium between atom and local environment can be investigated
by means of normal-mode quantum electrodynamics (QED)
\cite{Nakajima97,Wu00} or macroscopic QED in absorbing and dispersing
media \cite{buhmann07,scheel08}. In this case, thermal excitation and
de-excitation processes lead to resonant contributions to the force.
As discussed in a recent Letter by two of the present authors
\cite{buhmann08} (cf.~similar findings reported in
Ref.~\cite{Gorza06}), these resonant forces produce a different force
from that obtained through the standard approach, a perturbative
expansion of Lifshitz' formula using the ground state polarizability
of the atom. Only when the atom is fully thermalized, i.e., when it is
in a superposition of energy states as given by the Boltzmann
distribution, do the two approaches yield the same result, and only
when the correct, thermal polarizability is employed. For most atoms
the resonant contribution is small because the respective excitation
energies are much larger than the thermal energies, hence the atom is
essentially always in its ground state.

For diatomic polar molecules such as LiH or YbF, however, whose lowest
rovibrational eigenstates are typically separated by energies that are
small on a thermal scale, the situation is changed, and the
thermalized CP force can differ drastically from the standard
`Lifshitz-like' expression. An investigation into these effects was
undertaken in Ref.~\cite{ellingsen09} where it was found that for YbF
outside a metallic half-space the fully thermalized CP force is
smaller than the non-resonant force alone by a factor of 870. These
results could be of importance for the trapping of Stark-decelerated
polar molecules \cite{meerakker08} near macroscopic bodies.

Equally interesting is the observation that even for an atom or a
molecule in its ground state the resonant part of the Casimir-Polder
force has a long-range and spatially oscillating contribution, due to
propagating modes \cite{Wu00,ellingsen09}. While this oscillatory
behavior dies out as the system thermalizes, the thermalization time
of a ground state molecule can be quite long, often several seconds
\cite{buhmann08b}. The oscillating propagating potential reported in
\cite{ellingsen09}, unfortunately, was found to be too small in
amplitude to be useful for guiding purposes, but nonetheless points to
interesting applications if a way could be found to enhance these
oscillations.

In the present article we investigate the use of a planar cavity to
enhance the amplitude of the potential oscillations. This geometry
has been discussed in detail in conjunction with excited atoms in a
zero temperature environment, where an oscillating, standing-wave
potential is known to occur
\cite{Barton70,Barton79,Barton87,Jhe91,Jhe91b,Hinds91,Hinds94}. 
For ground-state molecules in a cavity at finite temperature, an
enhancement of up to two orders of magnitude will indeed be shown to
be possible when the cavity width is fixed at the resonant length
$a=\pi \hbar c/E_{10}$ where $E_{10}$ is the energy separation of the
ground state and first excited state of the molecule. 

The paper is organized as follows: The general formalism of the CP
force on a molecule in a cavity is developed in
Sec.~\ref{sec_formalism}, and numerical calculations for a gold
cavity are undertaken in Sec.~\ref{sec_enhancement}, where we also
show how the potential is enhanced as the cavity approaches the
resonant width. In Secs.~\ref{sec:resonance} and \ref{sec:species} we
investigate strategies for further enhancing the potential by
considering different cavity resonances and molecular species. The
scaling of the potential with the reflectivity of the cavity is
investigated numerically and analytically in Sec.~\ref{sec_scaling},
and we thereafter discuss how further enhancement can be achieved
using a cavity of parallel Bragg mirrors tuned to frequency
$\omega_{01}=E_{01}/\hbar$ and normal incidence (Sec.~\ref{Bragg}).
Finally, in Sec.~\ref{sec:lifetime} we investigate the effect of the
cavity on the thermalization time of a molecule initially prepared in
its ground state and find that this remains in the same order of
magnitude as in free space, typically in the order of seconds. We
summarize our result in Sec.~\ref{sec:conclusions} and provide a guide
to further investigations.

\section{Thermal Casimir-Polder potential in a planar cavity}
\label{sec_formalism}

We consider a polar molecule with energy eigenstates $|n\rangle$,
eigenenergies $\hbar\omega_n$, transition frequencies
$\omega_{mn}=\omega_m-\omega_n$ and dipole matrix elements
$\vect{d}_{mn}$, which is prepared in an incoherent superposition of
its energy eigenstates with probabilities $p_n$. As shown in
Ref.~\cite{buhmann08}, the CP force is conservative in the
perturbative limit, $\vect{F}(\vect{r}) = -\bm{\nabla}U(\vect{r})$,
where the associated CP potential is given by
\be
 U(\vect{r})=-\sum_{n}p_n U_n(\vect{r}),
\ee
and the potential components for an isotropic molecule read
\begin{align}\label{eq:thermoCP}
U_n(\vect{r})=&\,\mu_0\kB T\msum \xi_j^2\alpha_n (\mi\xi_j) \re\Tr
 \dyad{G}^{(1)}(\vect{r},\vect{r},\mi\xi_j)\notag\\&
 +\frac{\mu_0}{3}\sum_k\omega_{nk}^2 |\vect{d}_{nk}|^2 
 \{\Theta(\omega_{kn})n(\omega_{kn})\notag\\&
 -\Theta(\omega_{nk})[n(\omega_{nk})+1]\}\re \Tr 
 \dyad{G}^{(1)}(\vect{r},\vect{r},|\omega_{nk}|).
\end{align}
where $\mu_0$ is the free-space permeability, $\kB$ is Boltzmann's
constant, $\xi_j = 2\pi j \kB T/\hbar$ is the $j$th Matsubara
frequency, and $\dyad{G}^{(1)}(\vect{r},\vect{r}',\omega)$ is the
scattering part of the classical Green tensor of the
geometry the molecule is placed in. The prime on the Matsubara sum
indicates that the $j=0$ term is to be taken with half weight. The
molecular polarizability is given by
\begin{equation}
\label{eq:polarizabilityiso}
\alpha_n(\omega)
=\lim_{\epsilon\to 0}
\frac{1}{3\hbar}\sum_k\biggl[
 \frac{|\vect{d}_{nk}|^2}{\omega+\omega_{kn}+\mi\epsilon}
-\frac{|\vect{d}_{nk}|^2}{\omega-\omega_{kn}+\mi\epsilon}
 \biggr].
\end{equation}
The photon number follows the Bose-Einstein distribution,
\be\label{n}
  n(\omega) = 
  \left[\exp\left(\frac{\hbar\omega}{\kB T}\right)-1\right]^{-1}.
\ee
The first sum in Eq.~(\ref{eq:thermoCP}) is the non-resonant force,
reminiscent of that obtained by a dilute-gas expansion of Lifshitz'
formula \cite{lifshitz55}. The second sum is the resonant contribution
to the force. We will see how it splits naturally into a propagating
plus an evanescent part.

We assume the molecule to be placed within an empty planar cavity
bounded by two identical plates of infinite lateral extension with
plane parallel surfaces, separated by a distance $a$. We choose the
coordinate system such that the cavity walls are normal to the $z$
axis at $z=\pm a/2$ ($z=0$ being the center of the cavity) and denote
directions in the $xy$ plane by the symbol $\perp$. The scattering
Green tensor of the system is well known (cf., e.g.,
Ref.~\cite{tomas95}), and the relevant diagonal elements inside the
cavity are given by
\begin{subequations}\label{G}
\begin{align}
\label{eq:Gxx}
G^{(1)}_{xx}(z,z; \omega, \kp)  
 =& -\frac{\mi c^2 \beta}{\omega^2}\,\frac{r_p}{D_p}\,e^{\mi \beta a}
 \cos 2\beta z,\\
\label{eq:Gyy}
G^{(1)}_{yy}(z,z; \omega, \kp) =&
 \frac{\mi}{\beta}\,\frac{r_s}{D_s}\,e^{\mi \beta a}
 \cos 2\beta z,\\
\label{eq:Gzz}
G^{(1)}_{zz}(z,z; \omega, \kp) =& 
 \frac{\mi c^2\kp^2}{\omega^2\beta}\,
 \frac{r_p}{D_p}\,e^{\mi\beta a}\cos 2\beta z,
\end{align} 
\end{subequations}
where we have performed a Weyl expansion,
\be
\dyad{G}^{(1)}(\vect{r},\vect{r}',\omega) 
 = \int \frac{d^2 \kp}{(2\pi)^2}\,
 \dyad{G}^{(1)}(z,z',\vkp,\omega)\me^{\mi
 \vkp \cdot (\vect{r}-\vect{r}')_\perp},
\ee
taken the coincidence limit $\vect{r}\to\vect{r}'$ and dropped all
position-independent terms (which give rise to an irrelevant
constant contribution to the CP potential). Here, $r_s, r_p$
are the reflection coefficients of the (identical) cavity walls for
$s,p$ polarized waves and we have defined
\begin{align}
  D_\sigma &= 1-r_\sigma^2 \me^{2\mi \beta a}, \label{eq:D}\\
  \beta &= \sqrt{\omega^2/c^2 - \kp^2}.
\end{align}
The square root is to be taken such that $\im\beta\geq 0$. When the
cavity walls are homogeneous, semi-infinite half-spaces of an electric
material of permittivity $\varepsilon(\omega)$, the reflection
coefficients can be written simply as
\begin{subequations}
\begin{align}
\label{rs}
  r_s =& \frac{\beta - \sqrt{\beta^2 +
(\varepsilon-1)\omega^2/c^2}}{\beta + \sqrt{\beta^2 +
(\varepsilon-1)\omega^2/c^2}}\,, \\
\label{rp}
  r_p =& \frac{\varepsilon\beta - \sqrt{\beta^2 +
(\varepsilon-1)\omega^2/c^2}}{\varepsilon\beta + \sqrt{\beta^2 +
(\varepsilon-1)\omega^2/c^2}}\,,
\end{align}
\end{subequations}
where again the square roots are chosen such that their imaginary part
is positive.

Adding Eqs.~(\ref{eq:Gxx})--(\ref{eq:Gzz}) and partially performing
the Fourier integral by introducing polar coordinates in the $xy$
plane, the trace of the Green tensor of the cavity reads
\begin{align}
\label{eq:TrG}
&\Tr \dyad{G}^{(1)}(\vect{r},\vect{r},\omega)\notag\\
&=\frac{1}{2\pi\mi}\int_0^\infty \frac{\kp \dif\kp }{\beta}
 \Biggl[2\,\frac{c^2\beta^2}{\omega^2}\,\frac{r_p}{D_p}
 -\sum_{\sigma=s,p}\frac{r_\sigma}{D_\sigma}\Biggr]
 \me^{\mi\beta a}\cos 2\beta z.
\end{align}
This result can be substituted into Eq.~(\ref{eq:thermoCP}) to obtain
the thermal CP potential of a molecule in an arbitrary incoherent
internal state. In the following, we will assume the molecule to be
prepared in its ground state, so that the thermal CP potential is
given by
\begin{align}\label{eq:thermoCPground}
U(\vect{r})=&\,\mu_0\kB T\msum \xi_j^2\alpha(\mi\xi_j) \re\Tr
 \dyad{G}^{(1)}(\vect{r},\vect{r},\mi\xi_j)\notag\\&
 +\frac{\mu_0}{3}\sum_{k\neq 0}\omega_{0k}^2 n(\omega_{k0})
 |\vect{d}_{0k}|^2\re \Tr 
 \dyad{G}^{(1)}(\vect{r},\vect{r},\omega_{k0}),
\end{align}
[$\alpha(\omega)\equiv\alpha_0(\omega)$, ground-state polarizability]
together with Eq.~(\ref{eq:TrG}). The first term is the non-resonant
part of the potential, it depends on the Green tensor taken at purely
imaginary frequencies. Since $\beta$ is purely imaginary in this
case, the Green tensor~(\ref{eq:TrG}) and hence the non-resonant
potential is non-oscillating as a function of
position. The second term in the CP potential is the resonant
contribution, which depends on the Green tensor taken at real
frequencies. The integral over $\kp$ in this case naturally splits
into a region $0\leq \kp < \omega_{nk}$ of propagating waves in which
$\beta$ is real and positive, and a region $\omega_{nk}\leq \kp$ of
evanescent waves in which $\beta$ is purely imaginary. The
contributions from propagating waves are oscillating as a function of
position due to the term $\cos 2\beta z$, while those from evanescent
waves are non-oscillating, just like the non-resonant part of the
potential. The total potential~(\ref{eq:thermoCPground}) can thus be
separated into non-resonant (first term), propagating (contributions
to second term with $0\leq \kp < \omega_{nk}$), and evanescent
components (contributions to second term with $\omega_{nk}\leq \kp$)
according to 
\be
\label{CPcomp}
  U(z) = \Unr(z) + \Up(z) + \Ue(z)
\ee

\begin{figure}[!t!]
\includegraphics[width=3.3in]{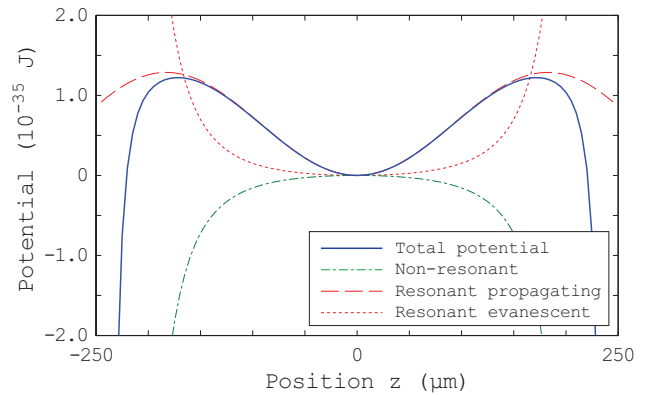}
\caption{Casimir-Polder potential of a ground-state LiH molecule
inside a gold cavity of width $a=500\mu$m at room temperature
($T=300$K). The non-resonant, propagating, and evanescent
contributions to the total potential are shown separately.}
\label{fig_components}
\end{figure}

To illustrate the behavior of the total potential and its three
components, we consider a LiH molecule in its electronic
and rovibrational ground state placed inside a gold cavity. The
permittivity of the (semi-infinite) cavity walls may be computed using
the Drude model
\be
  \varepsilon(\omega) = 1 - \frac{\wpsq}{\omega(\omega + \mi\gamma)}
\ee
with $\omega_\mathrm{p} = 1.37\times 10^{16}$rad/s and
$\gamma=5.32\times10^{13}$rad/s \cite{lambrecht00}. As shown in
Ref.~\cite{ellingsen09}, the CP potential of ground-state LiH is
dominated by contributions from the rotational transitions to the
first excited manifold, with the respective transition frequency and
dipole matrix elements being given by
$\omega_{0k}=2.78973\times10^{12}$rad/s and $\vdoisq=3.847\times
10^{-58}$ C$^2$m$^2$, respectively \cite{buhmann08b}. The
potential~(\ref{eq:thermoCPground}) and its three
components~(\ref{CPcomp}) for a cavity of length $a=500\mu$m at 
room temperature ($T=300$K) is shown in Fig.~\ref{fig_components} as
the result of a numerical integration, where
Eqs.~(\ref{eq:D})--(\ref{eq:TrG}) have been used. For transparency,
we have shifted all three components such that they vanish at the
center of the cavity.
It is seen that the non-resonant potential is attractive and has a
maximum at the center of the cavity, while the evanescent potential
is repulsive and has a minimum at the cavity center. As in the case
of a single surface \cite{ellingsen09} these two contributions
partially cancel, where the attractive non-resonant contribution is
slightly larger and leads to an attractive total potential in the
vicinity of the cavity walls. The propagating part of the potential
is spatially oscillating and finite at the cavity walls, it dominates
in the central region of the cavity where it gives rise to
well-pronounced maxima and minima.

It is natural to wonder whether these potential minima might be used
for the purpose of guiding of polar molecules. With this in mind, we
will in the following discuss strategies of enhancing the depth of the
potential well by analyzing the dependence of the potential on the
molecular species as well as the geometric and material parameters of
the cavity. 
 
\subsection{Cavity-induced enhancement of the
potential}\label{sec_enhancement}

We begin our analysis by discussing the dependence of the potential
on the cavity width. The one-dimensional confinement of the
propagating modes in a cavity of highly reflecting mirrors leads to
the formation of standing waves and associated cavity resonances. When
the molecular transition frequency coincides with one of these
resonances, the thermal CP potential can be strongly enhanced: When
the squared reflection coefficient $r_\sigma^2$ is close to unity, the
denominator $D_\sigma$ of Eq.~(\ref{eq:D}), featuring in the Green
tensor, becomes small if the exponential $\exp(2\mi \beta a)$ is
equal to unity, resulting in a strong enhancement of the potential
$\Up$. This happens for normal incidence ($\kp=0$) of the propagating
waves, when the resonance condition 
$2\omega_{0k}a/c = 2\pi\nu,~~\nu\in\mathbb{N}$ is
fulfilled. In other words, the cavity length has to be equal to a 
half-integer multiple of the molecular transition wavelength
$\lambda_{k0}=2\pi c/\omega_{k0}$:
\be
  a = \nu\lambda_{k0}/2,~~ \nu\in\mathbb{N}.
\ee
We say that the molecular transition coincides with the $\nu$th
cavity resonance. 

\begin{figure}[!t!]
\includegraphics[width=3.4in]{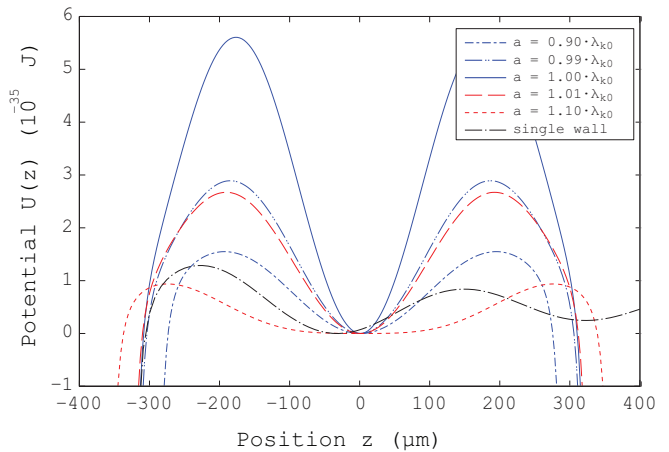}
\caption{
Cavity-induced enhancement of the thermal Casimir-Polder potential
of a ground-state LiH molecule inside gold cavities of various widths
close to the second resonance $a=\lambda_{k0}=673\mu\mathrm{m}$. The
potential of a single plate at $z=-\lambda_{k0}/2$ is also displayed.}
\label{fig_resonance}
\end{figure}

The cavity-induced enhancement of the thermal CP potential is
illustrated in Fig.~\ref{fig_resonance}, where we show the total
thermal CP potential of a ground-state LiH molecule in gold cavities of
widths such that the molecular transition is close to the second
cavity resonance $\lambda_{k0}$.
As seen, the amplitude of the spatial oscillations, associated with
the propagating part of the potential $\Up$, sharply increases as
the cavity width approaches $\lambda_{k0}$. For comparison, we have
also displayed the potential of a single plate at $z=-\lambda_{k0}/2$,
where Eq.~(\ref{eq:TrG}) for the cavity Green tensor has been replaced
with the single-plate result \cite{ellingsen09}
\begin{align}
\label{eq:TrGsingle}
&\Tr \dyad{G}^{(1)}(\vect{r},\vect{r}',\omega)\notag\\
&=\frac{\mi}{4\pi}\int_0^\infty \frac{\kp \dif\kp }{\beta}
 \sum_{\sigma=s,p}\biggl[r_\sigma
 -2\,\frac{c^2\beta^2}{\omega^2}\,r_p\biggr]
 \me^{\mi\beta (a+2z)}.
\end{align}
The comparison shows that the amplitude of the oscillations, while
hardly visible for the single plate, is strongly enhanced for a
cavity. The depth of the potential minimum at the center of the
cavity with respect to the neighboring maxima is increased by a factor
6.7 when using a resonant cavity rather than a single plate.

\subsection{Different cavity resonances}\label{sec:resonance}

\begin{figure}[!t!]
\includegraphics[width=2.8in]{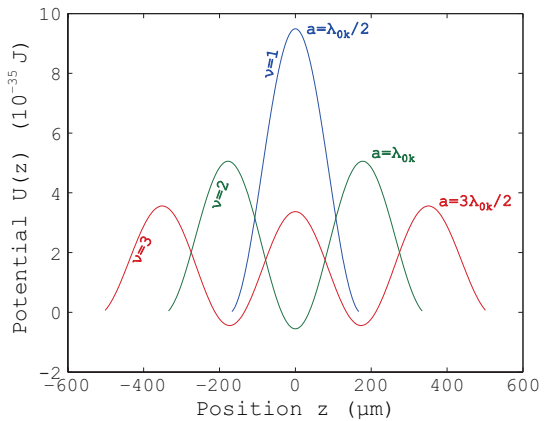}
\caption{
Propagating part of the thermal Casimir-Polder potentials of a
ground-state LiH molecule inside gold cavities of widths
$a=\nu\lambda_{k0}/2$ corresponding to resonances of different orders
($\nu=1,2,3$).
}
\label{fig_modeorder}
\end{figure}

In the following, we are interested in the cavity-enhanced
oscillations of the thermal potential. As seen from
Fig.~\ref{fig_resonance}, they set in at some distance away from the
cavity walls where the potential is well approximated by its
propagating-wave contribution $\Up$. We can therefore restrict our
attention to this part of the  total CP potential. The
(propagating-wave) potentials associated with different cavity
resonances $\nu$ are shown in Fig.~\ref{fig_modeorder}. 
It is seen that the order $\nu$ of the resonance corresponds to the
number of maxima of the potential. Potentials associated with
resonances of order $\nu\geq 2$ have minima. The amplitudes of the
oscillations become generally smaller for higher resonance orders
$\nu$. As seen from the case $\nu=3$, the minima and maxima are
slightly more pronounced towards the cavity walls.
 
The scaling of the potential minima with the resonance order as
observed in Fig.~\ref{fig_modeorder} can be confirmed by an analytical
analysis. For each cavity order $\nu$, we define $\Delta U_\nu$
to be the depth of the deepest potential minimum with respect to the
neighboring maxima. As suggested by Fig.~\ref{fig_modeorder}, this
deepest minimum will always be the one closest to the cavity walls. 
Cavity QED problems can often be solved analytically under the
simplifying assumption that reflection coefficients are independent of
the transverse wave number $k_\perp$ \cite{ellingsen08,ellingsen08e},
and this method is also successful here. As shown in App.~\ref{AppA},
in the perfect conductor limit $r_p=-r_s\equiv r\to 1$, we have 
the simple scaling law
\be\label{higherres}
 \Delta U_\nu  
 \propto\frac{1}{\nu}\,.
\ee
For imperfect conductors, the $\Delta U_\nu $ will decrease somewhat
less slowly with $\nu$.

The analytical scaling law obtained on the basis of simplifying
assumptions supports the observation from the numerical results in
Fig.~\ref{fig_modeorder} that the $\nu=2$ resonance provides the
deepest potential minimum. In view of potential guiding, we can
therefore restrict our attention to this case, $\Delta
U\equiv\Delta U_2$.

\subsection{Different molecular species}
\label{sec:species}

\begin{figure}[!t!]
\includegraphics[width=3.4in]{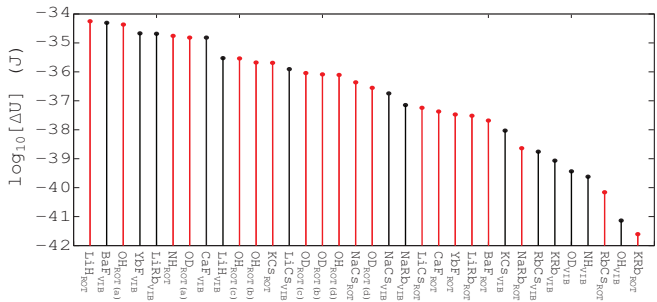}
\caption{Depth $\Delta U$ of the $\nu=2$ potential minimum for
rotational and vibrational transitions of various ground-state polar
molecules inside gold cavities at $T=300\mathrm{K}$. The different
non-degenerate transitions of OH and OD are labeled as (a)-(d) in
order of ascending frequencies, cf.~Ref.~\cite{buhmann08b}.
}
  \label{fig_variablemol}
\end{figure}

The CP potential depends on the molecular transition in question via
the respective transition frequencies and dipole matrix elements.
Using the molecular data as listed in Ref.~\cite{buhmann08b}, we have
calculated the depth of the $\nu=2$ potential minimum for both
rotational and vibrational transitions of the polar molecules LiH,
NH, OH, OD, CaF, BaF, YbF, LiRb, NaRb, KRb, LiCs, NaCs, KCs, and RbCs;
the results are displayed in order of descending $\Delta U$ in
Fig.~\ref{fig_variablemol}.
The figure shows that the deepest potential minima are realized when
using the rotational transition of LiH, the vibrational transition of
BaF or the dominant rotational transition of OH, followed by YbF
(vibrational), LiRb (vibrational), NH (rotational), OD (dominant
rotational transition), and CaF (vibrational).

The variation in the depth for different molecules is partly due to
its dependence on the molecular transition frequency. As shown in
Ref.~\cite{ellingsen09}, the resonant part of the CP potential of a
single plate is proportional to $\woi^2n(\woi)$ for a good conductor
with frequency-independent reflectivities. This remains true in the
case of a cavity. In addition, the amplitude of the oscillations is
inversely proportional to the molecule-wall separation. The largest
potential maximum being situated at $z-a/2=\lambda_{k0}/4\propto
1/\woi$, its height carries an additional $\woi$-proportionality. The
dependence of the potential-minimum depth on molecular transition
frequency can thus be given as
\be\label{freqdep}
 \Delta U_\nu  
 \propto\woi^3n(\woi)\propto
 \begin{cases}
 \woi^2\quad\mbox{for }\hbar\woi\ll\kB T,\\
 \me^{-\hbar\woi/\kB T}\quad\mbox{for }\hbar\woi\gg\kB T\,.
 \end{cases}
\ee
As shown in Sec.~\ref{sec_scaling}, this scaling becomes exact for
cavities with frequency- and $\kp$-independent reflectivities. For
real conductors, the decrease of $\Delta U_\nu$ for high frequencies
will be stronger than given in Eq.~(\ref{freqdep}) due to the
decrease of the reflection coefficients. Note that
Eq.~(\ref{freqdep}) also shows that $\Delta U_\nu$ becomes larger for
higher temperatures due to the increased thermal-photon number.
Again, this only holds when disregarding the temperature dependence
of the reflection coefficients, cf.\ also Sec.~\ref{Bragg} below.

\begin{figure}[!t!]
\includegraphics[width=3in]{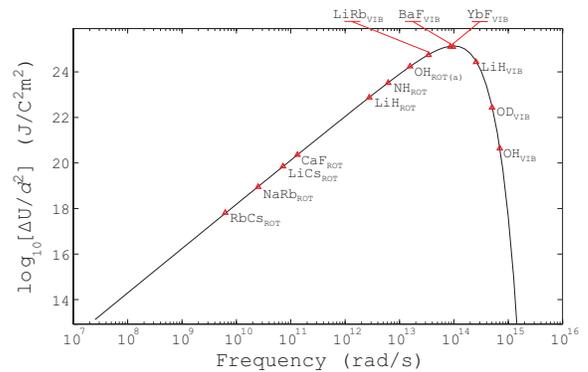}
\caption{Frequency-dependence of the depth $\Delta U$ of the $\nu=2$
potential minimum for polar molecules inside gold cavities at
$T=300\mathrm{K}$.}
  \label{fig_variableF}
\end{figure}

The frequency-dependence of $\Delta U_\nu$ is illustrated in
Fig.~\ref{fig_variableF} where we have plotted its values normalized
by dividing by the transition dipole moments $d^2$
($d^2\equiv\vdoisq$). 
The transition frequencies of some of the molecules investigated are
indicated in the figure. In particular, the vibrational transitions of
BaF and YbF, which have been seen in Fig.~\ref{fig_variablemol} to
give rise to large potential-minimum depths, are very close to the
peak of the function $\woi^3n(\woi)$, which is at
$\woi=1.11\times10^{14}\mathrm{rad}/\mathrm{s}$ for room
temperature.

The other main dependence of $\Delta U_\nu$ on the molecular species
and transition is the proportionality to the modulus squared of the
transition-dipole moments, 
\be\label{ddep}
 \Delta U_\nu  
 \propto\vdoisq=d^2.
\ee
The transition-dipole moments are typically larger for rotational
transitions than for vibrational ones. For this reason, the
rotational transition of LiH gives rise to the largest minimum depth
although the vibrational transition frequencies of BaF and YbF are
much closer to the peak frequency
$1.11\times10^{14}\mathrm{rad}/\mathrm{s}$.

\subsection{Scaling with reflectivity}\label{sec_scaling}

The cavity-induced enhancement of the thermal CP force strongly
depends on the reflectivity of the cavity walls. To understand this
dependence in more detail, let us for simplicity investigate how the
height of the single maximum for a $\nu=1$ resonance depends on
reflectivity. The scaling of the potential extrema with reflectivity
is the same for all $\nu$ as is shown in App.~\ref{AppA}, so
considering the simplest case will suffice.

We begin by writing the propagating part of the resonant CP potential
associated with a single transition in the form
\be
 \label{Ucalc}
 \Up(z) = \frac{1}{3\varepsilon_0}n(\woi)|\vect{d}_{01}|^2
 I(\phi)
\ee
where we have introduced the dimensionless position 
\be
  \phi = \frac{z}{a}
\ee
and the integral
\begin{multline}
\label{eq:I}
I(\phi) = \im\int_0^{{\woi}/c}\frac{d\kp\kp}{2\pi\beta}\Biggl[
 2\beta^2\frac{r_p}{D_p}
-\frac{\woi^2}{c^2}\sum_{\sigma=s,p}\frac{r_\sigma}{D_\sigma}
 \Biggr]\\ \times \me^{\mi\beta a}\cos 2\beta
 a\phi.
\end{multline}

As in Sec.~\ref{sec:resonance}, we consider the simple model case of
frequency- and $\kp$-independent reflection coefficients
$r_p=-r_s\equiv r$. With this assumption,
\be
  \sum_{\sigma=s,p}\frac{r_\sigma}{D_\sigma}=0.
\ee
After introducing the dimensionless integration variable $x=2\beta a$
with $\kp\dif\kp=-4a^2 x dx$, the integral above takes the form
\be
  I(\phi) = \frac{r}{8\pi a^3}\,\im\int_0^{x_0} \dif x\, 
 \frac{x^2e^{\mi x/2}\cos\phi x}{1-r^2 e^{\mi x}}
\ee
where $x_0=2\woi a/c$. For the $\nu=1$ resonance, we have
$a=\lok/2=\pi c/\woi$, so $x_0=2\pi$. 

The required height of the potential maximum at the cavity center
($z=0$) with respect to the value of $\Up$ at the cavity walls $z=\pm
a/2$ is proportional to the difference $I(0)-I(\frac{1}{2})$. We have
\begin{align}
  I(1/2) 
 =&\;\frac{r}{16\pi a^3}\,\im\int_0^{2\pi}\dif xx^2\,
 \frac{e^{\mi x}+1}{1-r^2\me^{\mi x}} \notag \\
  =&\;\frac{r}{16\pi a^3}\,\im\int_0^{2\pi} dxx^2\left[ 1
 +(1+r^{-2})\Li_0(r^2 e^{\mi x})\right] \label{inthalf}
\end{align}
where the polylogarithmic function is defined as
\be
  \Li_s(z) = \sum_{k=1}^\infty\frac{z^k}{k^s}.
\ee
The first term in Eq.~(\ref{inthalf}) is real and does not contribute. The
second one is easily calculated using the relation
\be
  \int dz \Li_s(A e^{b z}) = \frac1{b}\Li_{s+1}(Ae^{bz}) + C
\ee
valid for arbitrary constants $A,b$ where $|A|<1$. Partially
integrating this relation twice and substituting the result for
$A=r^2$, $b=\mi$ into Eq.~(\ref{inthalf}), one finds
\begin{align}
   I(1/2) =& \frac{r+r^{-1}}{16\pi a^3}\,
\im\left\{\frac{4\pi^2}{\mi}\Li_1(r^2)+ 4\pi\Li_2(r^2)
\right\}\notag \\
   =& \frac{\pi(r+r^{-1})}{4 a^3}\,\ln(1-r^2),\label{IhalfExact}
\end{align}
where we have noted that $\Li_1(z)=-\ln(1-z)$. In the limit of high
reflectivity, $\delta\equiv 1-r\to 0_+$ this exact result shows the
asymptotic behavior
\be
  I(1/2) \sim \frac{\pi}{2 a^3}(\ln\delta + \ln 2)
 \quad\mbox{for }\delta\to 0_+,
 \label{I1/2}
\ee
with the first correction term being of order $\delta$.

The calculation of $I(0)$ is only slightly more involved. We have
\begin{align}
I(0) =& \frac{r}{8\pi a^3}\,\im\int_0^{2\pi}\dif x 
 \frac{x^2\me^{\mi x/2}}{1-r^2\me^{\mi x}}\notag \\
  =&\frac{r}{8\pi a^3}\,\sum_{l=0}^\infty r^{2k}\im\int_0^{2\pi} dx
 x^2\me^{\mi x(l+\frac{1}{2})}.
\end{align}
By partial integration we obtain
\begin{equation}
\im\int_0^{2\pi}\dif x\, x^2\me^{\mi x(l+\frac{1}{2})} 
=\frac{4\pi^2}{(l+\frac{1}{2})}-\frac{4}{(l+\frac{1}{2})^3}.
\label{Inought}
\end{equation}
After substitution of this result, the sum over $l$ can be performed
by using the relations (cf.~\S1.513 in Ref.~\cite{BookGradshteyn80})
\be\label{lnrel}
\sum_{l=0}^\infty \frac{r^{2l}}{l+\frac1{2}} 
 = \frac{1}{r}\ln\frac{1+r}{1-r} 
 \sim - \ln\delta + \ln 2
 \quad\mbox{for }\delta\to 0_+
\ee
(leading corrections being of the order $\delta\ln\delta$) and
\begin{multline}
  \sum_{l=0}^\infty \frac{r^{2l}}{(l+\frac{1}{2})^3}
  \notag \\
  \sim8\left[\sum_{l=1}^\infty \frac{1}{l^3}
 -\sum_{l=1}^\infty\frac{1}{(2l)^3}\right] = 7\zeta(3)
 \quad\mbox{for }\delta\to 0_+,
\end{multline}
where $\zeta(z)$ is the Riemann zeta function. We thus find
\be
  I(0) \sim -\frac{\pi}{2 a^3}[\ln \delta - \ln 2 
  + \frac{7}{\pi^2} \zeta(3)] 
 \quad\mbox{for }\delta\to 0_+,
 \label{I0}
\ee
with the first correction again being of order $\delta\ln\delta$.

Substituting the results~(\ref{I1/2}) and (\ref{I0}) into
Eq.~(\ref{Ucalc}), the difference between the maximum and minimum
values of the $\nu=1$ propagating potential reads
\begin{align}
\label{Ur}
&\Up(0)-\Up(a/2) \sim -\frac{\pi\vdoisq n(\omega_{k0})}
 {3\varepsilon_0 a^3}
 \left[\ln\delta\!+\!\frac{7\zeta(3)}{2\pi^2} \right]\nonumber\\
&\quad\mbox{for }\delta\to 0_+.
\end{align}
This result being representative of the case of arbitrary $\nu$, we
can conclude that
\be\label{rdep}
 \Delta U_\nu  
 \propto\ln(1-r)
\ee
in the limit $r\to 1$.
In the case where the reflection coefficients are not the same for
both polarizations but still assumed constant, the coefficient of the
term $\propto \ln\delta$ in Eq.~(\ref{Ur}) will change, leading to a
slight quantitative but no qualitative difference to the scaling of
the potential-minimum depth. Note that Eq.~(\ref{Ur}) immediately
implies the scaling law~(\ref{freqdep}) for the frequency dependence
of $\Delta U_\nu$.  

The fact that the potential depth diverges only logarithmically as
reflectivity tends to unity poses severe restrictions on the potential
which is obtainable using a planar cavity. The mathematical reason
for the relative weakness of the resonance is that the integrand of
the $\kp$-integral only becomes large at a single point, at $\kp=0$.
The physical reason is that the photonic modes in the cavity are only
confined in one out of three spatial dimensions. We conjecture that
the potential due to the resonant CP force on a ground state molecule
can be much increased by a resonant cavity if confinement is imposed
in two or even three dimensions, i.e.\ in a cylindrical or spherical
cavity. 

The logarithmic scaling law of the potential-minimum depth 
$\Delta U_\nu$ for the case $\nu=2$ is confirmed by a numerical
calculation in which reflection coefficients are set constant,
$r_p=-r_s\equiv r$ and close to unity. 
\begin{figure}[!t!]
\includegraphics[width=3.3in]{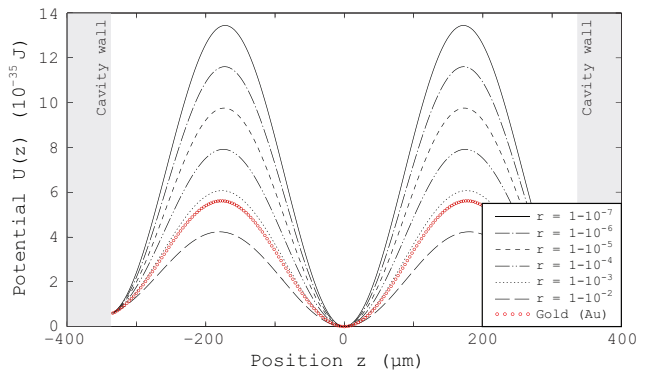}
\caption{Propagating part of the thermal CP potential for a
ground-state LiH molecule inside different cavities with constant
reflection coefficients. The rotational transition of LiH is assumed
to coincide with the $\nu=2$ resonance of the cavities. For
comparison, the exact result for a gold cavity is also shown.}
\label{fig_reflcoeff}
\end{figure}
The result for the rotational transition of LiH is shown in
Fig.~\ref{fig_reflcoeff} where the exact result for a gold cavity is
also included. By comparing the latter curve to the potentials for
constant reflection coefficients, one can read off the relatively
small `effective' reflectivity of gold between
$1-10^{-2}$ and $1-10^{-3}$ at the respective transition frequency of
LiH. For a molecule with a smaller eigenfrequency $\woi$ the gold
cavity does slightly better because the permittivity is larger.
Consider the vibrational transition of YbF with 
$\woi\approx 9\cdot 10^{10}$rad/s as an example, for which the
`effective' reflectivity of the gold cavity (in the sense of figure
\ref{fig_reflcoeff}) increases to about $1-10^{-3.5}$.

\subsection{Enhanced reflectivity using Bragg mirrors}\label{Bragg}

In contrast with the non-resonant CP force which depends on a very
broad band of frequencies, the resonant part of the ground state force
on a two-level molecule depends on the reflection properties of the
cavity at a single frequency, $\omega=\woi$. In addition, the
resonance of the cavity is also associated with a single value of
the wave vector $\vkp$, namely normal incidence. An enhancement of
the propagating potential hence does not require a good conductor like
gold which is a good reflector for a broad range of frequencies and
all angles of incidence; instead, cavity walls whose reflectivity has
a sharp peak at normal incidence and the single frequency $\woi$ are
sufficient. The obvious candidate is to use multilayer Bragg
mirrors, which consist of alternating layers of two different
materials, each layer of thickness being equal to one quarter of the
wavelength $\lambda_{10}=2\pi/n\woi$ in that layer where $n$ is the
respective refractive index.

The reflection coefficient of a stack of layers with permittivities 
$\varepsilon_j$ and thicknesses $d_j$ is found by recursive use of the
formula
\be
 r_{ijk\cdots} = 
 \frac{r_{ij}+r_{jk(l\cdots)}\me^{2i\beta_j d_j}}
 {1+r_{ij}r_{jk(l\cdots)}\me^{2i\beta_j d_j}}
\ee
($\beta_j=\sqrt{n_j^2\omega^2/c^2 - \kp^2}$), which relates the
reflection coefficient of a set of three adjacent layers $ijk\cdots$
(and all the layers behind) to the respective result for the next set
of adjacent layers $jkl\cdots$. If the $k$th layer is the last one of
the stack, the coefficients $r_{jk(l\cdots)}$ reduce to the two-layer
coefficients $r_{jk}$. In straightforward generalization of
Eqs.~(\ref{rs}) and (\ref{rp}), the two-layer coefficients read 
\begin{subequations}
\begin{align}
  r_{ij}^s=& \frac{\beta_i - \beta_j}{\beta_i + \beta_j}\,; \\
  r_{ij}^p=& \frac{\varepsilon_j\beta_i-\varepsilon_i\beta_j}
  {\varepsilon_j\beta_i+\varepsilon_i\beta_j}\;,
\end{align}
\end{subequations}
for $s$- and $p$-polarized waves, respectively. The Casimir effect for
such multilayer stacks has been extensively studied in the
past \cite{esquivel-sirvent01,raabe03,tomas05,ellingsen07}.

\begin{figure}[!t!]
\includegraphics[width=3.3in]{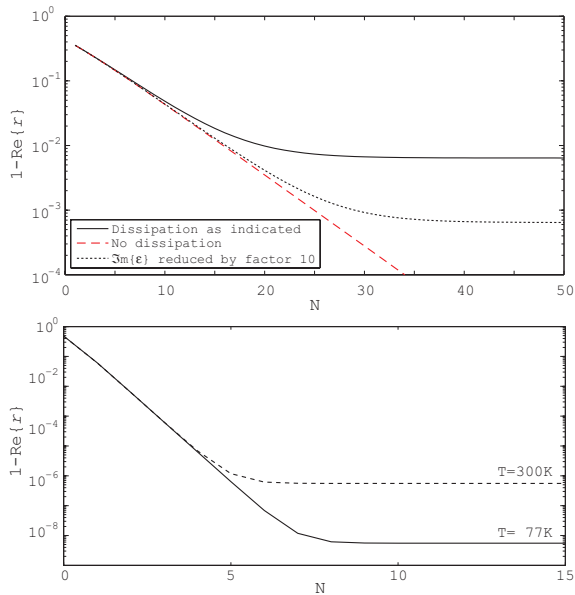}
\caption{Reflection coefficients of Bragg mirrors for normal
incidence at the rotational transition frequency of LiH \textit{vs}
the number of double layers $N$. \textit{Above:} GaAs/AlAs mirror,
where the results for reduced and vanishing absorption are also
displayed for comparison. \textit{Below:} Vacuum/sapphire mirror at
two different temperatures $T=77$K and $300$K.}
 \label{fig_BraggR}
\end{figure}

A very common pair of materials to use for Bragg mirrors is GaAs and
AlAs. At the rotational transition frequency of LiH, the permittivity
of the two materials can be roughly given as
$\varepsilon_\text{GaAs}=12.96 + 0.02\mi$ \cite{johnson69,courtney77}
and $\varepsilon_\text{AlAs}=10.96 + 0.02\mi$ \cite{adachi85}. The
reflection coefficient of a GaAs/AlAs Bragg mirror is plotted as a
function of the number of (double) layers $N$ in the upper panel of
Fig.~\ref{fig_BraggR}. For a given $N$, the Bragg mirror consists of
$2N+1$ layers in total, i.e. $N$ pairs of GaAs and AlAs layers of
thickness $\lambda_{10}/4$ (beginning with GaAs) and a terminating
GaAs layer of infinite thickness. 
As Fig.~\ref{fig_BraggR} shows, the reflectivity initially increases
for increasing $N$ and then eventually saturates for $N\gtrsim 30$ to
some finite value where $1-\re\,r\simeq 10^{-2}$. This saturation
is due to absorption, as is illustrated by the other two curves, where
we have given the results that would be obtained for a reduced or
vanishing imaginary part of the permittivities. For a reduced
imaginary part, the saturation sets in for higher $N$, and
consequently to a lower $\delta$. In the absence of absorption, the
reflectivity could be brought arbitrarily close to unity by adding
more and more layers.

A higher reflectivity could hence be obtained by using materials with
very small dielectric loss. One example of such a Bragg mirror could
be alternating layers of vacuum and sapphire, which can have an
extremely low loss tangent ($\im\,\varepsilon/\re\,\varepsilon\simeq
10^{-5}$ and $10^{-7}$ at room temperature and $77$K, respectively
\cite{driscoll92}) combined with a refractive index considerably
larger than unity ($\re\,\varepsilon\simeq 10$ \cite{wang08}). Using
the approximative values $\varepsilon_\text{sapph}=10+10^{-4}\mi$ at
$300$K and $\varepsilon_\text{sapph}=10+10^{-6}\mi$ at $77$K, we have
computed the reflection coefficients of the vacuum/sapphire mirror as
displayed in the lower panel of Fig.~\ref{fig_BraggR}. At room
temperature, the coefficient saturates at $N\gtrsim 6$ to 
$\delta =5.5\cdot 10^{-6}$. At $T=77$K, the reflection coefficient
saturates at $N\gtrsim 8$ to $\delta = 5.5\cdot 10^{-8}$, the the
increase in reflectivity is obviously due to the reduction of
material absorption for the lower temperature. Note that in
comparison to the GaAs/AlAs mirror, the number of layers required for
saturation is significantly lower because of the larger dielectric
contrast; and the room-temperature reflectivity at saturation is
increased by about four orders of magnitude.

\begin{figure}[!t!]
\includegraphics[width=3.4in]{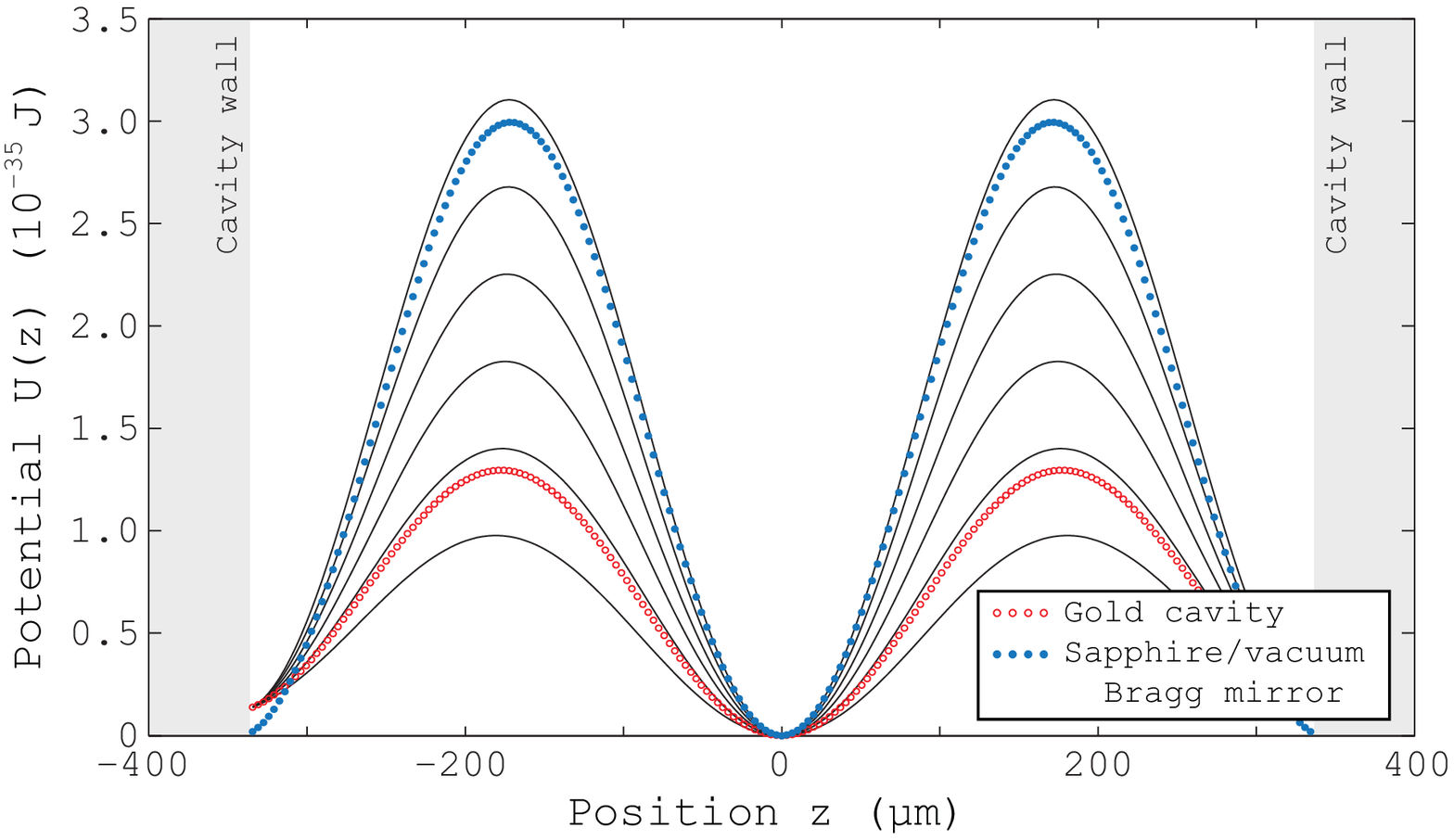}\\
\includegraphics[width=3.4in]{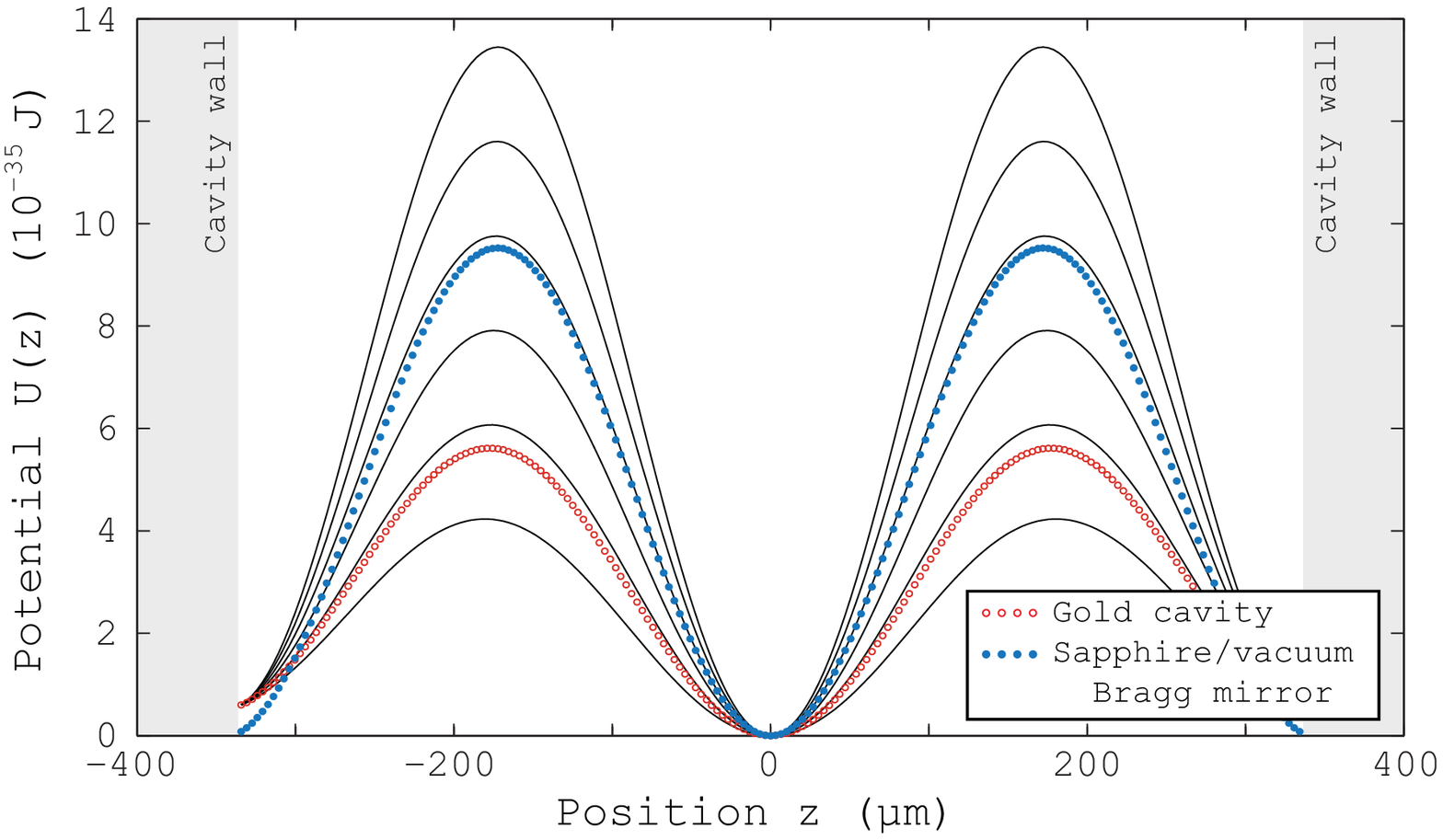}
\caption{Resonant part of the thermal CP potential associated with
the rotational transitions of a ground-state LiH molecule at $\nu=2$
resonance with a gold cavity and a cavity bounded by vacuum/sapphire
Bragg mirrors at two different temperatures: $77$K above, $300$K
below. The solid black lines represent calculations at constant
reflection coefficients as in Fig.~\ref{fig_reflcoeff}; the
corresponding values of $1-r$ decrease in powers of $10$ from
$10^{-2}$ (lowest curve) to $10^{-7}$ (highest curve). The same
permittivity is used for gold for both temperatures.}
\label{fig_BraggCP}
\end{figure}
The resulting propagating part of the resonant CP potential at
resonant cavity width using the sapphire/vac\-uum Bragg mirror at
$T=77$K and $300$K are shown in Fig.~\ref{fig_BraggCP}, where the
corresponding graphs at various constant reflection coefficients have
also been displayed for reference. The `effective' reflection
coefficients achieved at the two temperatures are around
$\delta=10^{-4.8}$ and $\delta = 10^{-6.7}$ respectively, and the
potential depths approximately a factor $2.45$ and $1.77$ greater than
that of the gold cavity at the same temperatures. Note, however, that
the effect of the enhanced reflectivity at $77$K is counteracted by
the overall decrease of the potential due to the lower photon number.

\subsection{Lifetime of the ground state in the cavity}
\label{sec:lifetime}

Resonant CP potentials are only present for molecules which are not at
equilibrium with their thermal environment, i.e., on a time scale
given by the inverse heating rate \cite{ellingsen09}. When enhancing
the thermal CP potential via a resonant cavity, it is necessary to
ascertain that the simultaneous cavity-enhancement of heating rates
does not reduce the lifetime of the resonant potential by so much as
to render it experimentally inaccessible. We show in the following
that the lifetime of the molecular ground state is not radically
changed even by the presence of a resonant planar cavity.

The total heating rate of an isotropic molecule out of its ground
state may be written as \cite{buhmann08}
$\Gamma = \Gamma_0 + \Gamma_\text{cav}$ where 
\be
  \Gamma_0 = \frac{\vdoisq\woi^3n(\woi)}{3\pi\hbar c^3 \varepsilon_0}
\ee
is the heating rate in free space and
\be
  \Gamma_\text{cav} = 
 \frac{2\mu_0}{3\hbar}\vdoisq\woi^2
 n(\woi) \im \Tr  \dyad{G}^{(1)}(\vect{r},\vect{r},\woi)
\ee
is its change due to the presence of the cavity. Apart from the
prefactor, this additional term has the same form as the expression
for the potential, except that the imaginary part of the Green tensor
is taken rather than the real part. 

\begin{figure}[!t!]
\includegraphics[width=3.4in]{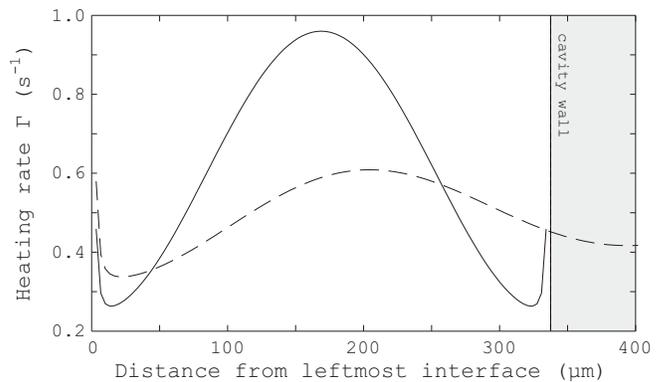}
\caption{Ground-state rotational heating rate of a LiH molecule in a
gold cavity at $\nu=1$ resonance $a=\pi c/\woi$ (solid line) and to
the right of a gold half-space (dashed line) at room temperature
($T=300$K).}
  \label{fig_lifetime}
\end{figure}

In Sec.~\ref{sec_scaling}, we had shown that for real and constant
reflection coefficients, the Green tensor exhibits a logarithmic
divergence as $r\to 1$ with a purely real coefficient, whereas all
other contributions remain finite. This shows that the imaginary part
of the Green tensor responsible for the decay rate can be expected to
remain finite even for strongly increased reflectivity. It
follows that the presence of the cavity does not drastically change
the lifetime of the ground state of the molecule, which will typically
be in the order of seconds. This is confirmed in
Fig.~\ref{fig_lifetime} where we display the ground-state heating rate
of a LiH molecule inside a $\nu=1$ gold cavity and near a gold
half-space. The lifetime is reduced by only a factor 2 at the center
of the cavity, remaining in the order of seconds.

\section{Conclusions and outlook}
\label{sec:conclusions}

We have studied the thermal Casimir-Polder potential of ground-state
polar molecules placed within a planar cavity at room temperature. As
was previously found in Ref.~\cite{ellingsen09}, the resonant
absorption of thermal photons by a molecule gives rise to spatial
oscillations of that potential. Our results demonstrate that the
amplitude of these oscillations is enhanced when placing the molecule
inside a suitable cavity such that a molecular transition frequency
coincides with a cavity resonance. We have analyzed the dependence of
this oscillating potential on the parameters of the molecule and the
cavity by both analytical and numerical means and found that the depth
of potential minima $\ldots$
\begin{itemize}
\item \emph{Cavity resonance:} $\ldots$ decreases with increasing
order of the cavity resonance approximately as $1/\nu$;
\item \emph{Molecular eigenfrequency:} $\ldots$ is proportional to   
$\woi^3n(\woi)$ for good conductors, where $n(\woi)$ is the thermal
photon number;
\item \emph{Molecular dipole moments:} $\ldots$ is proportional to the
modulus squared $\vdoisq=d^2$ of the total transition dipole moment; 
\item \emph{Temperature}: $\ldots$ increases with temperature due to
an increase of the thermal photon number $n(\woi)$;
\item \emph{Reflectivity of cavity walls}: $\ldots$ scales as 
$\ln(1-r)$ for high reflectivity $r$.
\end{itemize}

In view of observing this potential and possibly utilizing it for
the guiding of cold polar molecules, these observations imply the
following strategies for enhancing the depth of the potential minima:
\begin{itemize}
\item \emph{Cavity resonance:} The $\nu=2$ resonance is most
suitable, since it gives the deepest minimum.
\item \emph{Molecular species:} At room temperature, the deepest
minima are realized for molecules whose transitions are not too far 
from the peak frequency $1.11\times 10^{14}\mathrm{rad}/\mathrm{s}$
and which at the same time feature suitably large transition dipole
moments. Good candidates are, e.g., LiH (rotational transitions), BaF
(vibrational transitions) or OH (rotational transitions).
\item \emph{Cavity walls:} Highly reflecting cavities are required in
order to enhance the potential. Bragg mirrors consisting of
materials with small absorption such as sapphire are favorable to
single layers of good conductors like gold. 
\item \emph{Temperature:} Temperatures should be in the range of room
temperature or even higher in order to achieve large photon numbers.
This should be balanced, however, against the adverse reduction of
reflectivity of most materials with increasing temperature.
\end{itemize}

With an optimum choice of all these parameters, the planar cavity can
be used to enhance the resonant potential by one or at most two orders
of magnitude with respect to the single-plate case. However, the thermal
potentials achievable with planar cavities are in all likelihood still
too small to facilitate the guiding of polar molecules. 

The limitations of the enhancement of the potential in a planar cavity
are ultimately due to the weak (logarithmic) scaling with
reflectivity. A stronger scaling may be expected in geometries
providing mode confinement in more than just one dimension such as
cylindrical or spherical cavities. This will be investigated in a
future publication. Note that apart from the different expected
scaling with reflectivity, all other conclusions regarding the
dependence of the potential on the relevant molecular and material
parameters as given above hold irrespective of the geometry under
consideration. The strategies for the enhancement of thermal CP
potentials developed in this work will thus present a valuable basis
when considering more complicated cavity geometries.


\acknowledgments
This work was supported by the Alexander von Humboldt Foundation,
the UK Engineering and Physical Sciences Research Council, and the
SCALA programme of the European Commission. S.\AA.E. acknowledges
financial support from the European Science Foundation under the
programme `New Trends and Applications of the Casimir Effect'.

\appendix
\section{Scaling of potential depth with resonance order}
\label{AppA}

For a cavity of width $a = \nu\lambda_{k0}/2$ the potential has $\nu$
peaks, roughly located at
$z=-\nu\lambda_{k0}/4+(\mu-1/2)\lambda_{k0}/2$
($\mu=1\ldots\nu$) and $\nu-1$ minima at
$z = -\nu\lambda_{k0}/4 + \mu\lambda_{k0}/2$ ($\mu=1\ldots\nu-1$).
For a given resonance order $\nu$, the deepest minima are the
ones closest to the cavity walls (and which have a maximum on both
sides), for example the rightmost one at $z = (\nu-2)\lambda_{k0}/4$.
It has to be compared with the lower of the two adjacent maxima, i.e.,
the one immediately to the right at $z=(\nu-3)\lambda_{k0}/4$. The
required depth of the deepest minimum is hence given by
\be\label{deltaUnu1}
 \Delta U_\nu  
 =U[(\nu-3)\lambda_{k0}/4]-U[(\nu-2)\lambda_{k0}/4].
\ee
We calculate this depth for a cavity whose reflection coefficients are
independent of the transverse wave number $k_\perp$, 
$r_p=-r_s\equiv r$, and close to unity, $\delta=1-r\ll 1$. Noting that
the oscillating part of the potential is determined by $\Up$ and
introducing the definition~(\ref{eq:I}), cf.~Sec.~\ref{sec_scaling},
we thus have
\be\label{deltaUnu}
 \Delta U_\nu  
 \propto I\bigl({\textstyle\frac{1}{2}-\frac{3}{2\nu}}\bigr)
 -I\bigl({\textstyle\frac{1}{2}-\frac{1}{\nu}}\bigr).
\ee

We consider in the following only the terms which do not vanish as
$\delta\to 0_+$. For arbitrary $\phi$, we have
\be
  I(\phi) = \frac{r}{2\pi\nu^3\lok^3}\im \int_0^{2\pi\nu}\dif x
  \frac{x^2 e^{ix/2}\cos\phi x}{1-r^2 e^{ix}}
\ee
which, after expanding the fraction in powers of $r^2$, solving the
integral over $x$ and taking the imaginary part, can be written as
\be
  I(\phi) =
\frac{r}{2\pi\nu^3\lok^3}\sum_{j=0}^\infty\left[y(j+\tfrac{1}{2}
+\phi)+y(j+\tfrac{1}{2}-\phi)\right]
\ee
with
\begin{align}
  y(p) =& -\frac{2}{p^3} +
\left(\frac{2}{p^3}-\frac{4\nu^2\pi^2}{p}\right)\cos(2\pi\nu p) \notag
\\
  &+ \frac{4\nu\pi}{p^2}\sin(2\pi\nu p).
\end{align}
In particular, this implies
\begin{subequations}\label{Iapp}
\begin{align}
  I(\textstyle\frac1{2}-\textstyle\frac{3}{2\nu}) =&
\frac{2r}{\pi\nu^3\lok^3}\sum_{j=0}^\infty
r^{2j}\left[\frac{\nu^2\pi^2}{j+1-\textstyle\frac{3}{2\nu}}+\frac{
\nu^2\pi^2}{j+\textstyle\frac{3}{2\nu}}\right.\notag \\
  & \left.-
\frac{1}{(j+1-\textstyle\frac{3}{2\nu})^3}-\frac{1}{(j+\textstyle\frac
{3}{2\nu})^3}\right],\\
  I(\textstyle\frac1{2}-\textstyle\frac{1}{\nu}) =&
-\frac{2r\pi}{\nu\lok^3}\sum_{j=0}^\infty
\left[\frac{r^{2j}}{j+1-\textstyle\frac{1}{\nu}}+\frac{r^{2j}}{
j+\textstyle\frac{1}{\nu}}\right]
\end{align}
\end{subequations}
The evaluation of sums with simple denominators can be performed by
using the relation (formula~9.559 in Ref.~\cite{BookGradshteyn80}) 
\be
  \sum_{j=0}^\infty \frac{r^{2j}}{j+b} = \frac1{b} F(1,b;1+b;r^2),
\ee
valid for any $b\neq 0,-1,-2,...$ Here, $F(a,b;c;z)\equiv {}_2
F_1(a,b;c;z)$ is a hypergeometric function which in turn has the
following expansion in powers of $\delta=1-r$ (formula~15.3.10 in
Ref.~\cite{BookAbramowitz64})
\begin{align}
  \frac1{b}F(1,b;1+b;r^2) 
  \sim& -\ln \delta -\ln 2-\gamma - \psi(b)
 \label{appEq}
\end{align}
as $\delta\to 0_+$, the correction terms being of order
$\delta\ln\delta$. Here, $\psi(x)$ is the logarithmic derivative of
the gamma function and $\psi(1)=-\gamma$ where $\gamma=0.577216$ is
Euler's constant. 

For the sums in Eq.~(\ref{Iapp}) with cubic denominators, one can set
$r=1$ with an error of order $\delta$. The sums are then simply
Hurwitz zeta functions $\zeta(3,b)$ 
\begin{align}
  \sum_{l=0}^\infty\frac1{(l+b)^3}\equiv& \zeta(3,b). 
\end{align}

\begin{figure}[!t!]
\includegraphics[width=2.9in]{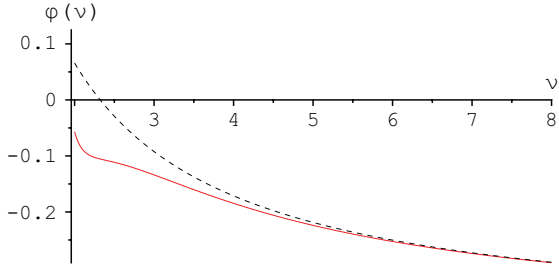}
\caption{Graph of $\varphi(\nu)/\nu$ (solid, red line) and its
asymptote (dashed line) as given by
Eq.~(\ref{asymptote}).}
\label{fig_app}
\end{figure}

We thus find
\be
  I\bigl({\textstyle\frac{1}{2}-\frac{3}{2\nu}}\bigr)
 -I\bigl({\textstyle\frac{1}{2}-\frac{1}{\nu}}\bigr) =
-\frac{8\pi}{\nu\lok^3}[\ln \delta + \varphi(\nu)] + ...
\ee
as $\delta\to 0_+$ with corrections being of order
$\delta\ln\delta$ and
\begin{align}
\label{eq:a11}
  \varphi(\nu) \equiv& \ln 2 + \gamma +
\textstyle\frac{1}{4}\psi(1-\textstyle\frac{3}{2\nu})+
\textstyle\frac{1}{4}\psi(\textstyle\frac{3}{2\nu})\notag \\
  &
+\textstyle\frac{1}{4}\psi(1-\textstyle\frac{1}{\nu})+\textstyle\frac{
1}{4}\psi(\textstyle\frac{1}{\nu})\notag \\
&+\frac{1}{4\pi^2\nu^2}\left[
\zeta(3,1-\textstyle\frac{3}{2\nu})
+\zeta(3,\textstyle\frac{3}{2\nu})\right].
\end{align}
Some numerical values of $\varphi(\nu)$ are
\begin{subequations}
\begin{align}
  \varphi(2)=&-0.1134423724; \\
  \varphi(3)=&-0.4015949503; \\
  \varphi(4) =& -0.7384479470. 
\end{align}
\end{subequations}
We give a plot of $\varphi(\nu)/\nu$ in Fig.~\ref{fig_app}.
For large $\nu$, one obviously has
\begin{align}
  \varphi(j) \sim& -(\textstyle\frac{5}{12}-\frac{2}{27\pi^2})\nu+
\ln(2)+\frac1{4}+...\notag \\
  \approx& -0.4091613938j + 0.9431471806 +...\label{asymptote}
\end{align}
plus terms of order $\nu^{-1}$ and smaller. Already at $\nu=4$ this is
a fairly good approximation to Eq.~(\ref{eq:a11}) as shown in
Fig.~\ref{fig_app}.


\end{document}